\documentclass[sigconf, nonacm]{acmart}

\usepackage{booktabs}
\usepackage{listings}
\usepackage{xcolor}
\usepackage{microtype}
\usepackage{tikz}
\usetikzlibrary{positioning,arrows.meta,fit,backgrounds,patterns,shapes.geometric,decorations.pathreplacing,calc}

\AtBeginDocument{%
  \providecommand\BibTeX{{%
    \normalfont B\kern-0.5em{\scshape i\kern-0.25em b}\kern-0.8em\TeX}}}

\setcopyright{none}

\title{How Much Coordination Gain Is Real? A Paired Noise-Floor Protocol for Multi-Agent LLM Benchmarks}

\author{Alibek T Kaliyev}
\email{alibek.kaliyev@utexas.edu}
\affiliation{%
  \institution{The University of Texas at Austin}
  \city{Austin}
  \state{Texas}
  \country{USA}
}

\author{Artem Maryanskyy}
\email{artem.maryanskyy@uber.com}
\affiliation{%
  \institution{Uber}
  \city{San Francisco}
  \state{California}
  \country{USA}
}

\begin{document}

\begin{abstract}
% Shorter, first-read-accessible abstract: problem -> measurement ->
% finding -> metric -> substrate. Detailed numbers moved to body.

Multi-agent LLM coordination papers report small
benchmark deltas as evidence that one architecture beats
another. A prior question: how much paired
trial-$0$ disagreement do two protocols produce on the same
model and benchmark when their API inputs are
\emph{configuration-equivalent} (matched by code inspection
plus a SHA-$256$ byte audit), short of full
identity-replay? On Claude Haiku $4.5$ against
$\tau^2$-bench retail, the clean configuration-equivalent
contrast (no\_coord vs.\ intercept, both inert at trial $0$)
gives signed paired gaps of $+10$\,pp and $0$\,pp across two
$n{=}100$ seeds; pooled across both, $+5$\,pp with Wilson CI
$[{-}2,{+}12]$, not significant. The largest single-seed
contrast ($+18$\,pp pull-vs-intercept, $p_{\text{corr}}{=}0.012$)
did not reproduce at the second seed ($-3$\,pp,
$p_{\text{corr}}{=}1.0$); no trial-$0$ contrast is
significant after Bonferroni at either seed or pooled. The envelope of
observed paired gaps spans $[{-}3,{+}18]$\,pp across two seeds,
with pooled upper Wilson CI $\lesssim 15$\,pp.
\textbf{Seven of ten recent multi-agent coordination
architectures report headline effects below this local floor,
and one more sits inside the envelope}; whether they survive a same-model paired
replication is, by construction, untested in their original
settings.

We define \emph{coordination-active pass$^k$}, pass$^k$
restricted to trials where the coordination mechanism is
logically active, as the minimum reporting protocol, with
sample-size targets and runtime hooks in the body. Measurements run on \textbf{ET-MCP}, a
task-scoped negative-knowledge store conformant with MCP
2026-07-28, used as a substrate to isolate reader-side
choices, not as a contribution. On Haiku $4.5$ the
candidate readers (pull, intercept) do not improve
trial-$1$ recovery; we give a preliminary diagnosis of
failure modes with refinements on existing production hook
surfaces~\cite{strandshooks2025,agentcorepolicy2025,langchainmiddleware2026,sagemaker_monitor2022}.

\end{abstract}

\maketitle

\section{Introduction}
\label{sec:introduction}

Two LLM agents work in parallel on the same multi-step itinerary.
The first attempts to book a flight, discovers no inventory,
abandons the path, and moves on. The second, started moments
later by the same orchestrator, attempts the same booking
through the same API and re-discovers the same dead end. The
first agent's failure lived only in its private scratchpad. This
is a structural property of how multi-agent LLM systems (MAS)
currently coordinate~\cite{mast2025,codedelegator2026}. Recent
architectures address it with coordination channels (structured
handoffs, shared memory, server-side state) and report small
benchmark gains as evidence one design beats another.

\paragraph{Why we wrote this paper.}
We initially set out to evaluate two such designs (an
agent-side \emph{pull} channel and a framework-side
\emph{intercept} channel) against a no-coordination baseline on
$\tau^2$-bench retail. The head-to-head shows no detectable
effect at the available power: the coordination-active subsets
contain $n{=}8$--$17$ informative pairs after ties, far short of
what a medium-sized improvement would require. Unpacking that
non-result forced a question we should have asked at the start:
\emph{what is the run-to-run variance between protocols whose
API inputs are equivalent at trial $0$?} A single number is
enough to invert the read of much of the recent coordination
literature. The rest of the paper answers that question and
traces what it implies for our own and others' architectural
claims.

\paragraph{ET-MCP as substrate.}
Measuring variance under matched payload required a
representative coordination substrate, so we built one.
\textbf{ET-MCP} is an Ephemeral Trace store of typed
negative-knowledge events exposed via the Model Context Protocol
(MCP) 2026-07-28~\cite{mcpspec2026}: task-scoped, append-mostly,
with a TF-IDF (term-frequency / inverse-document-frequency)
ranker over $\textsc{failed\_path}$-typed events. Two reader-side
variants ride on the same store. \emph{Pull} makes the agent the
active reader: it calls \texttt{trace.query} when peer state
might inform the next decision. \emph{Intercept} makes the
framework the active reader: it consults the store transparently
on every tool call and prepends a \texttt{[PEER-WARNING]} block
when (name, arguments) match a peer event. Intercept rides on
the same hook surface already deployed for blocking, validation,
retries, and caching in
Strands~\cite{strandshooks2025}, AgentCore
Policy~\cite{agentcorepolicy2025},
LangChain~\cite{langchainmiddleware2026}, and the
Microsoft Agent Framework~\cite{msagentframework2026}.

\paragraph{The result that reframed the paper.}
Running this substrate against its no-coordination baseline
yielded no detectable effect at the available power. At
$\tau^2$-bench retail, Haiku $4.5$, $T{=}0$, $n{=}100$ tasks,
pull, intercept, and no\_coord all reach trial-$1$ success
$0.54$ (paired sign tests fail to reject, $p{=}1.0$ for all
three contrasts;
\S\ref{sec:eval:results:retail-intercept}).
Decomposed by trial, no\_coord and intercept are
\emph{configuration-equivalent} at trial $0$: their requests to
the API are equivalent by code inspection (system prompt, tool
list, sampling parameters, message-array construction; see
\S\ref{sec:eval:request-equivalence} for what we did and did not
verify). This is the only genuinely clean contrast in the
trio; the paired sign-test disagreement is $21/11/68$ at
seed-$1$ and $13/13/74$ at seed-$2$ ($+10$\,pp and $0$\,pp
signed gaps; pooled $+5$\,pp, Wilson CI $[{-}2,{+}12]$,
Bonferroni $p_{\text{corr}}{=}0.711$, not significant). The
largest single-seed paired contrast was pull-vs-intercept at
seed-$1$ ($27/9/64$, $+18$\,pp, $p_{\text{corr}}{=}0.012$),
but it did not reproduce at seed-$2$ ($18/21/61$, $-3$\,pp,
$p_{\text{corr}}{=}1.0$); the pooled gap is
$+7.5$\,pp with Wilson upper CI ${\sim}15$\,pp. We characterize
the configuration-equivalent envelope on this benchmark, model,
and harness as a paired-gap distribution whose observed range
across two seeds is $[{-}3,{+}18]$\,pp and whose pooled upper
Wilson CI is $\lesssim 15$\,pp---no single-seed Bonferroni
significance survives a second seed. The floor is local,
not universal; cross-model and cross-domain probes (Haiku
airline $n{=}30$, Sonnet retail $n{=}30$) confirm it does not
transfer.

\paragraph{Coordination-active pass$^k$.}
To separate run-to-run variance from mechanism effect we define
$\textsc{pass}^k_{\textrm{active}}$: marginal pass$^k$
restricted to tasks where the peer-coordination store was
non-empty at trial start. Marginal pass$^k$ answers ``should I
deploy the whole system?'' Active pass$^k$ answers ``did
coordination help when it had an opportunity to.'' The two
diverge whenever empty-store trials carry unrelated
configuration perturbations, which is the situation we encounter.
Three runtime alarms operationalize the metric on Amazon Bedrock
AgentCore and SageMaker Model
Monitor~\cite{sagemaker_monitor2022}
(\S\ref{sec:eval:deployment}).

\paragraph{Contributions.}
Four contributions track the threads above:
\begin{itemize}
    \item \textbf{(C1) Measurement protocol.} A same-model,
        paired, configuration-equivalent variance procedure for
        coordination-architecture claims on state-validated MAS
        benchmarks. The floor splits into an
        \emph{identity-replay} component (byte-identical
        payloads, which we do not measure) and a
        \emph{configuration-equivalent} component (same intended
        setup, code-inspection-verified, which we
        do)~(\S\ref{sec:eval:request-equivalence}).

    \item \textbf{(C2) Finding.} On $\tau^2$-bench retail with
        Haiku $4.5$ across two $n{=}100$ seeds, the clean
        configuration-equivalent paired-gap pooled CI is
        $[{-}2,{+}12]$\,pp (no\_coord vs.\ intercept);
        no single-seed contrast is significant after Bonferroni
        once a second seed is measured. The observed envelope
        spans $[{-}3,{+}18]$\,pp with pooled upper Wilson CI
        $\lesssim 15$\,pp. Seven of ten recent multi-agent
        coordination architectures
        (Table~\ref{tab:literature-vs-floor})
        report headline gains below this envelope and one more
        sits inside it; each falls below or inside the envelope
        measured here.

    \item \textbf{(C3) Metric.}
        $\textsc{pass}^k_{\textrm{active}}$ (coordination-active
        pass$^k$): pass$^k$ conditioned on a non-empty
        coordination store. Sample-size guidance and three
        runtime instrumentation alarms ship the metric as a
        pre-deployment release gate
        (\S\ref{sec:eval:deployment}).

    \item \textbf{(C4) Failure-mode diagnosis (preliminary).}
        The trial-level data surface three candidate
        failure modes of naive negative-knowledge coordination
        (writer mis-attribution; per-match injection noise;
        brittle keying), with matched architectural refinements
        on existing production hook surfaces
        (\S\ref{sec:proposals}). Single-judge audit and small-$n$
        validation leave these as hypotheses for confirmatory work.
\end{itemize}

ET-MCP itself and the harness mitigating a documented
litellm~\cite{litellmissue12404,litellmissue15322} bug class
appear as supporting artifacts rather than claimed contributions.

\section{Background and Motivation}
\label{sec:background}

\paragraph{Coordination failure is measured, not anecdotal.}
The Multi-Agent System Taxonomy (MAST)~\cite{mast2025}
attributes $36.94\%$ of failures in $1{,}600+$ production agent
traces to inter-agent coordination breakdowns, the single
largest category, and CodeDelegator~\cite{codedelegator2026}
documents context pollution scaling with handoff payload size.
Coordination is therefore a measurable target, not an
anecdotal one; the reliability of those measurements is
the load-bearing question.

\paragraph{ET-MCP as measurement substrate.}
We use ET-MCP, a task-scoped negative-event store exposed
through MCP 2026-07-28~\cite{mcpspec2026} primitives, as the
substrate under test. The architectural details are secondary
to this paper's claim; what matters is that ET-MCP is an
MCP-conformant coordination layer that can be toggled on and
off in paired runs, isolating a coordination-active arm against
a matched baseline. The remainder of the paper is about what
that paired comparison reveals about variance, not about ET-MCP
itself.

\section{Related Work}
\label{sec:related}

\paragraph{Prior coordination systems.}
CA-MCP~\cite{camcp2026} is the nearest prior system by name and
substrate: server-to-server, positive plan state, single-trial;
the authors explicitly disclaim cross-trial use.
Terrarium~\cite{terrarium2025} is a second MCP-based system
with a pull-style interface, but targets attack-surface analysis.
MPAC~\cite{mpac2026} proposes Lamport-clock conflict resolution
for the multi-principal case. Reflexion and
ExpeL~\cite{reflexion2023,expel2024} share the negative-knowledge
intuition but inject via prompt context and are single-agent or
cross-task-transfer. Long-lived cross-task memory
(MemGPT~\cite{memgpt2024}, G-Memory~\cite{gmemory2025}) operates
at task-spanning rather than within-task granularity.
AgentVerse and related orchestration substrates surveyed
in~\cite{mast2025} round out the prior-systems comparison
revisited in Table~\ref{tab:literature-vs-floor}. Framework-level hooks
(Strands~\cite{strandshooks2025}, AgentCore
Policy~\cite{agentcorepolicy2025},
LangChain middleware~\cite{langchainmiddleware2026},
Microsoft Agent Framework~\cite{msagentframework2026},
AutoGen~\cite{autogen2023}, MetaGPT~\cite{metagpt2024})
carry blocking, validation, retries, and caching but no
coordination semantics, and any of them could host the same
measurement protocol.

\paragraph{Measurement methodology: the actual gap.}
A concurrent thread measures the reliability of agent
benchmarks. Yu~et~al.~\cite{multiagentmem2026} frame the
multi-agent case as lacking remote direct memory access.
Variance-decomposition work spans a 12-metric
framework~\cite{rabanser2026reliability}, an ICC
framework~\cite{mustahsan2025stochasticity}, Anthropic-side
infrastructure noise~\cite{anthropicnoise2026}, aggregate
pass@1 variance~\cite{randomnessagentic2026}, systematic
agent reliability~\cite{reliabilitybench2026}, and input-noise
robustness~\cite{agentnoisebench2026}.
\textbf{None of these reports paired configuration-equivalent
replication on a state-validated multi-agent coordination
benchmark, and none conditions on a coordination-active
sub-event.} Bedi~et~al.~\cite{bedi2025paradox} (KDD~'25,
same workshop series) document a component-vs-system paradox
in clinical multi-agent settings but treat the system as a
black box. Cuadron~et~al.'s
$\tau^2$-bench-verified~\cite{tau2bench_verified2025} fixes
tasks while ours fixes transport; the two compose. AgentTrace's
post-hoc causal graphs~\cite{agenttrace2026} and
STATE-Bench~\cite{statebench2026}'s memory-agnostic harness
are complementary rather than competitive.
The central claim of this paper, that coordination effect
sizes should be reported relative to a same-model paired
coordination-active variance gate, is what fills that gap.
Ou~et~al.~\cite{ou2025coordination} compare
orchestrator-mediated vs.\ shared-notebook coordination on
TravelPlanner but neither pair at matched payload nor report
a coordination-active variance floor, and we revisit them in
Table~\ref{tab:literature-vs-floor} as one of the recent
coordination papers whose reported effects straddle the paired
noise envelope measured here on Haiku~4.5 / $\tau^2$-bench
retail (pooled upper CI $\lesssim 15$\,pp).

\section{ET-MCP Architecture}
\label{sec:architecture}

ET-MCP is the substrate that makes the measurement protocol of
\S\ref{sec:evaluation} possible; it is not the contribution of
this paper. This section covers only what the measurement story
needs: the minimal tool surface, the failure-only writer policy,
and the pull-vs-intercept reader-side distinction that
\S\ref{sec:evaluation} compares.

\subsection{Tool surface and lifecycle}
\label{sec:arch:tools}
\label{sec:arch:lifecycle-write-query}

The substrate is structured around one trace namespace per
\texttt{task\_id}, served over the stateless MCP 2026-07-28
transport~\cite{mcpspec2026}. Agents see three operations:
\texttt{trace.write} (append a typed event),
\texttt{trace.query} (TF-IDF ranking over the typed payload,
returning top-$k$ peer events with one-line summaries), and
\texttt{trace.cas} (compare-and-swap for the rare double-commit
race). The task lifecycle (\texttt{init}, \texttt{complete}, TTL
sweep) is owned by the orchestrator through a separate channel
that participating agents cannot reach, so no agent can tear down
a peer's namespace. The TF-IDF ranker is a single replaceable
component.

\subsection{Writer policy}
\label{sec:arch:taxonomy}

Events carry an envelope (\texttt{event\_id}, \texttt{task\_id},
\texttt{event\_type}, \texttt{agent\_id}, \texttt{timestamp},
\texttt{version}, \texttt{payload}). Five typed event categories
cover the negative-outcome cases mapped to the MAST failure
taxonomy~\cite{mast2025}: failed paths, constraint violations,
abandoned approaches, irreversible intermediate decisions, and
tool errors. The default \texttt{failure\_only} client-side
selection policy writes only on these negative outcomes. The
$\tau^2$ retail evaluation specializes this to a single
\textsc{failed\_trial\_action} type keyed on the trial's terminal
reward.

\subsection{Two reader-side architectures: pull and intercept}
\label{sec:arch:pull-vs-intercept}

\begin{figure*}[t]
\centering
% Side-by-side architecture diagram comparing pull and intercept.
% Two-column figure (figure* in 04-architecture.tex).
% Semantic color encoding consistent across panels:
%   blue   = store read (curated summary / consult)
%   gray   = ordinary agent <-> tool I/O
%   red    = framework-injected / peer-warning payload (intercept only)
% Step labels are numbered per panel (P1..P3 / I1..I5); the caption
% decodes every step in prose so the figure itself stays uncluttered.
% Requires: \usetikzlibrary{positioning,arrows.meta,fit,backgrounds,shapes.geometric,calc}
\begin{tikzpicture}[
  font=\small,
  agent/.style={draw, rectangle, rounded corners=2pt,
                minimum width=1.7cm, minimum height=0.7cm,
                inner sep=2pt, fill=blue!10, align=center},
  tool/.style={draw, rectangle, rounded corners=2pt,
               minimum width=1.7cm, minimum height=0.7cm,
               inner sep=2pt, fill=green!10, align=center},
  framework/.style={draw, rectangle, rounded corners=2pt, dashed, thick,
                    minimum width=2.4cm, minimum height=0.7cm,
                    inner sep=2pt, fill=orange!12, align=center},
  store/.style={draw, cylinder, shape border rotate=90, aspect=0.25,
                minimum width=2.0cm, minimum height=0.8cm,
                inner sep=2pt, fill=gray!12, align=center, font=\footnotesize},
  flow/.style={-{Stealth[length=2mm,width=1.8mm]}, thick, draw=black!65, line width=0.6pt},
  flowread/.style={-{Stealth[length=2mm,width=1.8mm]}, thick, draw=blue!75!black, line width=0.75pt},
  flowwarn/.style={-{Stealth[length=2mm,width=1.8mm]}, thick, draw=red!75!black, line width=0.85pt, dashed},
  stepnum/.style={font=\bfseries\scriptsize, circle, draw=black!70, inner sep=0.4pt,
               minimum size=3.6mm, fill=white},
  stepwarn/.style={font=\bfseries\scriptsize, circle, draw=red!70!black, inner sep=0.4pt,
                   minimum size=3.6mm, fill=red!8},
  stepread/.style={font=\bfseries\scriptsize, circle, draw=blue!70!black, inner sep=0.4pt,
                   minimum size=3.6mm, fill=blue!6},
  panel/.style={draw=black!60, thick, rounded corners=4pt, inner sep=10pt, fill=white}
]

% =========================================================
% LEFT PANEL: PULL  (agent-as-reader)
% Layout: store top, agent + tool side-by-side below.
% =========================================================
\begin{scope}[local bounding box=pullbox]
  \node[store]                    (s1) at (0, 2.2) {ET-MCP store};
  \node[agent, below=14mm of s1]  (a1)              {Agent};
  \node[tool, right=22mm of a1]   (t1)              {Tool};

  % (P1) store -> agent: curated summary at trial start
  \draw[flowread] (s1.south) -- (a1.north)
    node[stepread, pos=0.45] {P1};

  % (P2) agent -> tool: tool call (upper lane)
  \draw[flow]
    ([yshift=3mm]a1.east) -- ([yshift=3mm]t1.west)
    node[stepnum, pos=0.5] {P2};

  % (P3) tool -> agent: raw response (lower lane)
  \draw[flow]
    ([yshift=-3mm]t1.west) -- ([yshift=-3mm]a1.east)
    node[stepnum, pos=0.5] {P3};
\end{scope}
\begin{scope}[on background layer]
\node[panel, fit=(pullbox),
      label={[font=\bfseries]above:Pull \emph{(agent-as-reader)}}] (pullpanel) {};
\end{scope}

% =========================================================
% RIGHT PANEL: INTERCEPT  (framework-as-reader)
% Layout: store top, framework below store, agent + tool side-by-side below framework.
% Curved arrows so request (top) and response (bottom) lanes are clearly separable.
% =========================================================
\begin{scope}[xshift=10cm, local bounding box=icbox]
  \node[store]                       (s2) at (0, 2.5) {ET-MCP store};
  \node[framework, below=8mm of s2]  (fw)             {Framework};
  \node[agent, below=22mm of fw, xshift=-22mm]   (a2) {Agent};
  \node[tool,  below=22mm of fw, xshift=22mm]    (t2) {Tool};

  % I4: framework consults store (vertical, straight)
  \draw[flowread] (fw.north) -- (s2.south)
    node[stepread, pos=0.5] {I4};

  % I1: agent -> framework (upper curve, bending right/up)
  \draw[flow, bend left=20]
    (a2.north) to
    node[stepnum, pos=0.55] {I1}
    ([xshift=-3mm]fw.south);
  % I5: framework -> agent (lower curve, bending right/down, red dashed)
  \draw[flowwarn, bend left=20]
    ([xshift=-6mm]fw.south) to
    node[stepwarn, pos=0.45] {I5}
    ([xshift=3mm]a2.north);

  % I2: framework -> tool (upper curve, bending left/down)
  \draw[flow, bend left=20]
    ([xshift=3mm]fw.south) to
    node[stepnum, pos=0.45] {I2}
    (t2.north);
  % I3: tool -> framework (lower curve, bending left/up)
  \draw[flow, bend left=20]
    ([xshift=-3mm]t2.north) to
    node[stepnum, pos=0.55] {I3}
    ([xshift=6mm]fw.south);
\end{scope}
\begin{scope}[on background layer]
\node[panel, fit=(icbox),
      label={[font=\bfseries]above:Intercept \emph{(framework-as-reader)}}] (icpanel) {};
\end{scope}

% Compact semantic legend below both panels (single line).
\node[font=\scriptsize, below=2.5mm of pullpanel.south west, anchor=north west,
      xshift=2mm] {%
  \tikz[baseline=-0.5ex]{\draw[flowread] (0,0)--(0.55,0);}\;store read\quad
  \tikz[baseline=-0.5ex]{\draw[flow] (0,0)--(0.55,0);}\;agent$\leftrightarrow$tool\quad
  \tikz[baseline=-0.5ex]{\draw[flowwarn] (0,0)--(0.55,0);}\;injected \texttt{[PEER-WARNING]}};

\end{tikzpicture}
\caption{Reader-side architecture contrast. The two designs share
the event store, writer policy, and MCP transport; they differ
only in who reads and when the payload is presented.
\textbf{Pull} (left): P1 store $\to$ agent (curated summary at
trial start); P2 agent $\to$ tool (call); P3 tool $\to$ agent
(raw response).
\textbf{Intercept} (right): I1 agent $\to$ framework (call); I2
framework $\to$ tool (forward); I3 tool $\to$ framework (raw);
I4 framework $\to$ store (consult); I5 framework $\to$ agent
(response, with \texttt{[PEER-WARNING]} prepended on argument
match).  Pull presents the curated payload once; intercept
injects raw warnings per matching call.}
\label{fig:pull-vs-intercept}
\Description{Side-by-side architecture diagram. Left panel shows
the pull architecture: the ET-MCP store sends a curated summary to
the agent at trial start, then the agent calls the tool and the
tool returns a raw response. Right panel shows the intercept
architecture: the agent calls the tool, the call passes through
the framework which consults the ET-MCP store, the tool returns
its raw response to the framework, and the framework forwards a
modified response with a [PEER-WARNING] block prepended back to
the agent.}
\end{figure*}
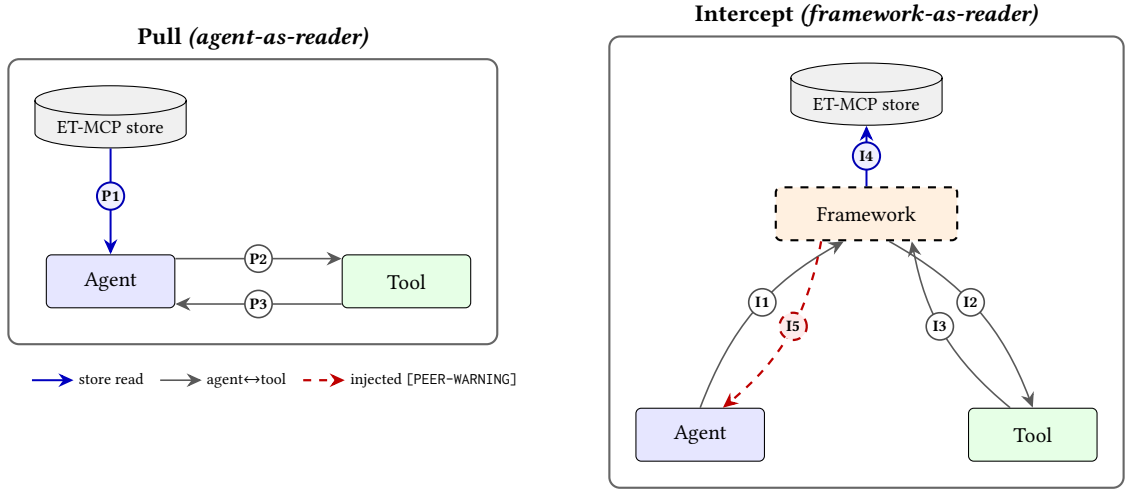

Figure~\ref{fig:pull-vs-intercept} contrasts the two designs.
\textbf{Pull} places the agent as the \emph{active reader}: it
calls \texttt{trace.query} when it estimates peer events might
inform the next decision, paying decision overhead, round-trip
latency, and an enlarged tool surface for that control.

\textbf{Intercept} relocates the reader into the framework. The
orchestrator intercepts every tool call, and if the call's
(name, arguments) appears in the task-scoped store as a peer
\textsc{failed\_path} or \textsc{failed\_trial\_action} event,
it prepends a \texttt{[PEER-WARNING]} block to the response.
Intercept removes the agent-side decision, the LLM round-trip,
and any agent code changes; the tool surface and prompt are
unchanged. The hook surface intercept rides on is already a
production primitive in Strands~\cite{strandshooks2025},
AgentCore Policy~\cite{agentcorepolicy2025}, LangChain
middleware~\cite{langchainmiddleware2026}, and Microsoft Agent
Framework~\cite{msagentframework2026}, so intercept is an
additive extension to existing runtimes, not a new framework.

Both architectures share the same event taxonomy, failure-only
writer policy, task-scoped lifecycle, and MCP transport, and are
configuration-equivalent at trial 0: pull's augmenter is a no-op
when the store is empty, and the intercept hook is pass-through
on no argument match. The full audit of this equivalence is in
\S\ref{sec:eval:request-equivalence}. The next
section instantiates both in the reference implementation that
produces the measurements of \S\ref{sec:evaluation}.

\section{Implementation}
\label{sec:implementation}

To produce clean measurements, we built an Anthropic-native
$\tau^2$-bench harness (\textasciitilde$700$ lines) that replaces
only the conversation-state construction layer; $\tau^2$'s domain
database, tools, task definitions, and DB-hash reward are
unchanged, and no model, sampling, or scoring code is touched, so
the harness rewrite cannot itself contribute to the trial-$0$
disagreement of \S\ref{sec:eval:request-equivalence}. This bypasses a documented bug class in the stock
$\tau^2$-bench/litellm
pipeline~\cite{litellmissue12404,litellmissue15322} that corrupts
$\geq 30\%$ of Haiku 4.5 trials, enabling zero-infrastructure-error
sweeps. The harness and the \texttt{et\_mcp} substrate of
\S\ref{sec:architecture} ship together in the open-source
repository.

\section{Evaluation}
\label{sec:evaluation}

\subsection{Benchmarks and conditions}
\label{sec:eval:benchmarks}
\label{sec:eval:conditions}
\label{sec:eval:ablations}

\textbf{$\tau^2$-bench retail}~\cite{tau2bench2025} is the
primary benchmark for the reader-side architecture head-to-head.
It scores state-validated pass$^k$ on a customer-service
tool-using domain.  We run $n{=}100$ tasks (seed 42), with Haiku
4.5 as both assistant and user simulator, a 20-turn budget, and
2 trials per task.  Cross-model and cross-domain probes
(\S\ref{sec:eval:results:sonnet}) use Sonnet 4.5 retail and
$\tau^2$-bench airline at $n{=}30$ each.

\subsection{Metrics and statistical methodology}
\label{sec:eval:metrics}
\label{sec:eval:stats}

\textbf{Completion rate} is the fraction of trials whose final
output passes the benchmark scorer.  \textbf{Pass$^k$} on
$\tau^2$-bench follows the unbiased estimator of Chen et al.: the
marginal success rate at $k{=}1$, and the both-trials-succeed
rate at $k{=}2$.  Two protocols are
\textbf{configuration-equivalent} at a given trial when code
inspection confirms identical request-side augmenter behavior at
that trial (system prompt prefix, tool-response modifications,
request headers, sampling parameters).  This is necessary but not
sufficient for byte-identical wire payloads
(\S\ref{sec:eval:request-equivalence}).

\textbf{Coordination-active pass$^k$
($\textsc{pass}^k_{\mathrm{active}}$)} restricts pass$^k$ to tasks
where the peer-coordination store is non-empty at each of trials
$1\!\ldots\!k$, the structurally-defined coordination-active
sub-event. It is causally pre-specifiable from the protocol
semantics: when the store is empty the coordination mechanism is
inactive, so any cross-protocol comparison on store-empty trials
measures noise rather than architecture. For $k{=}2$ in our
setting this is equivalently the trial-1 success rate on tasks
where trial~0 failed. Each contrast uses its own paired subset
(tasks where both compared protocols failed trial~0), so per-row
rates are not jointly comparable across contrasts.

Effect size throughout is Cliff's $\delta$ with the Romano et
al.\ (2006) magnitude thresholds. We apply paired sign tests
(binary pass$^k$ on $\tau^2$; the equivalent McNemar test agrees
and is reported in Table~\ref{tab:mde}). The $\tau^2$
head-to-head and coordination-active contrasts are reported
outside the Bonferroni correction families as the pre-specified
primary tests; all multiplicity corrections in this paper are
Bonferroni with $m{=}3$. All trials are logged to JSONL artifacts under
\texttt{eval/results/}; analysis is reproducible via the
\texttt{analysis} CLI in the repository.

\paragraph{Pre-registration.}  The Haiku $n{=}100$ retail
head-to-head is the pre-specified primary test: protocols,
$K{=}3$ writer, seed $42$, and $T{=}0$ were frozen before the run
($K{=}3$ was pilot-tuned on a prior $n{=}30$
sweep---\S\ref{sec:eval:limits}).
$\textsc{pass}^k_{\mathrm{active}}$ is causally pre-specifiable but
was applied to already-frozen data.  Sonnet, airline, M1, and P2
are exploratory; Sonnet $n{\geq}250$ retail is pre-registered as
confirmatory.

%==================================================================
% RESULTS
%==================================================================

\subsection{Results}
\label{sec:eval:results}

\subsubsection{Request-equivalence: what the trial-$0$ floor
measures}
\label{sec:eval:request-equivalence}

The trial-$0$ paired gap decomposes into four layered sources;
naming them keeps the headline numbers honest. The first is a
byte-identical first-request audit (E1); the second is
within-protocol API stochasticity at $\leq 3$\,pp (E2); the
third is between-protocol harness drift once the user-simulator
emits its first sampled output (E3); the fourth is a
hypothesized pull-specific structural perturbation (E4), which
the second seed did not confirm.
Strictly speaking, what we measure at trial $0$ is multi-turn
paired variance between two configuration-equivalent harness
runs; we label it a \emph{coordination noise floor} because it
is the gate any coordination claim on this benchmark must
clear, not because the variance itself is coordination-specific.

At trial~$0$ the store is empty, so reader hooks are no-ops. The
three arms share the agent system prompt
(\texttt{AGENT\_SYSTEM\_TEMPLATE}; pull's augmenter returns input
unchanged when \texttt{warnings\_block()} is empty), the
protocol-independent tool list (no \texttt{trace.query}
injection), user-simulator prompt and priming, sampling parameters
(\texttt{T}{=}$0$, \texttt{max\_tokens}{=}$4096$, same
\texttt{claude-haiku-4-5}), and a pass-through intercept hook.
Two SHA-$256$ audits confirm E1: an offline reconstruction of
\texttt{client.messages.create} kwargs ($40/40$ cells
byte-identical: $n{=}10$ tasks $\times$ $4$ first-request
payload components, each hashed across all $3$ protocols) and a
live wire-byte audit ($30/30$ user-simulator opens
byte-identical on $10$ headline tasks).

The Messages API exposes no \texttt{seed}, so $T{=}0$ requests
are not bit-deterministic; a paired $n{=}30$ within-protocol
replicate (L3) bounds E2 at $\leq 3$\,pp. Once the user-simulator
emits its first sampled output, it becomes the agent's first
message and downstream wire bytes diverge across runs (within
\emph{or} between protocol); that drift is E3. Pull's
seed-$1$ $p_{\text{corr}}{=}0.012$ initially looked like a
small real structural perturbation in pull's augmenter (E4);
the second seed did not reproduce it (\S\ref{sec:eval:results:retail-intercept}),
so E4 is now a candidate-not-confirmed source rather than an
established one. E1--E3 together (plus E4 as a hypothesis) form
the \emph{configuration-equivalent} envelope, and the
coordination-active metric is designed to condition away E2--E4.
The head-to-head below measures any remaining effect against it.

\subsubsection{Reader-side architecture head-to-head: pull vs.\
intercept on $\tau^2$-bench retail}
\label{sec:eval:results:retail-intercept}

\textit{What this shows.} On the coordination-active subset
(trial~1 of tasks where trial~0 failed), pull and intercept tie
and both directionally underperform the no-coordination
baseline; with $n{=}8$--$17$ informative pairs after ties, the
subset is underpowered for a medium-sized effect, so we read
this as no detectable effect at this power rather than a clean
null. The trial-$0$ paired gaps anchor an empirical noise floor
that any architectural claim must clear. Pooled across two
$n{=}100$ seeds, the clean configuration-equivalent contrast
(no\_coord vs.\ intercept) gives $+5$\,pp (CI $[{-}2,{+}12]$,
not significant); the largest contrast (pull vs.\ intercept)
pools to $+7.5$\,pp with upper Wilson CI ${\approx}15$\,pp; and
the largest single-seed gap ($+18$\,pp, seed-$1$,
$p_{\text{corr}}{=}0.012$) did not reproduce at seed-$2$.

$\tau^2$-bench retail~\cite{tau2bench2025} reports state-validated
pass$^k$ on customer-service conversations
(\S\ref{sec:eval:benchmarks}). All runs use the
\textasciitilde$700$-line Anthropic-native harness
(\S\ref{sec:implementation}). A pilot $n{=}30$ smoke on the stock
$\tau^2$-bench/litellm pipeline observed $\geq30\%$ of Haiku 4.5
trials terminating with malformed
\texttt{tool\_use}/\texttt{tool\_result} sequences, silent at the
harness layer, consistent with the documented bug
class~\cite{litellmissue12404,litellmissue15322}. Our harness
observed zero infra errors across $600$ trials.

\paragraph{Protocols.}
The three protocols share an identical writer side.  At the close
of any trial with reward $<\!1$, the final $k{=}3$ tool calls
are written as \textsc{failed\_trial\_action} events in the
task-scoped store.  The $\tau^2$ harness uses this simplified
single event type instead of the full five-category taxonomy
of \S\ref{sec:arch:taxonomy} because retail trials
rarely surface explicit tool errors mid-trial; the failure signal
is the trial's terminal reward.  ($k{=}3$ is a causal-attribution
heuristic, pilot-tuned on a prior $n{=}30$ sweep; the $n{=}100$
run confirms this configuration rather than pre-registering
$k{=}3$ as optimal.)

The three protocols differ only in the reader interface.
\texttt{no\_coord} leaves the store unused. \texttt{pull}
(agent-as-reader) prepends a curated peer-warnings summary to the
agent's system prompt at trial start. \texttt{intercept}
(framework-as-reader) prepends a \texttt{[PEER-WARNING]} block to
the tool \emph{response} when the call's (name, arguments) match
a \textsc{failed\_trial\_action} event. Any pass$^k$ difference
between pull and intercept under matched payload is therefore
attributable to the reader-side architecture, and we observe none
at trial~1.

\begin{figure*}[t]
\centering
\includegraphics[width=0.84\textwidth]{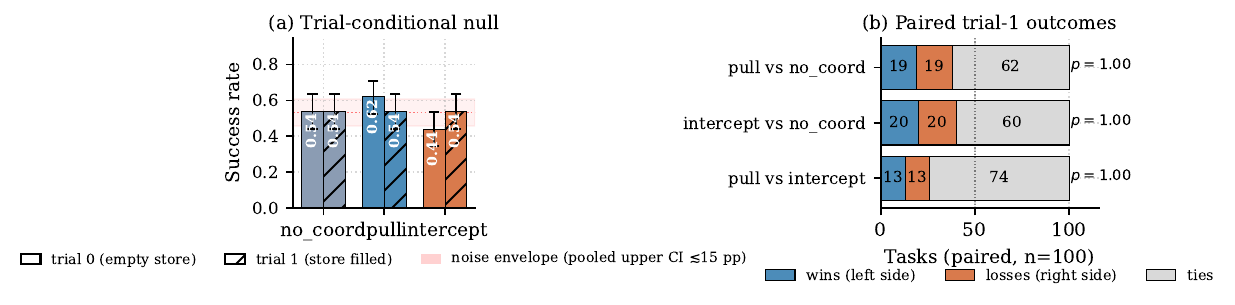}
\caption{Headline visual (seed-$1$ of two $n{=}100$ seeds;
\S\ref{sec:eval:results:retail-intercept} reports the second
seed and pooled tests). The coordination-active head-to-head
shows no detectable effect at the available power; the trial-$0$
paired gap is the noise-floor estimate.
\textbf{(a) Coordination-active head-to-head.}  Success rate by
protocol at trial~0 (empty store, coordination inert) and
trial~1 (store filled).  \texttt{no\_coord} and intercept are
configuration-equivalent at trial~0 by code inspection; their
$+10$\,pp signed paired gap at this seed (Bonferroni
$p_{\text{corr}}{=}0.330$, CI $[{-}2,{+}16]$ crosses zero)
pools to $+5$\,pp (CI $[{-}2,{+}12]$) across both seeds and is
the clean empirical noise floor. Pull's $+18$\,pp paired gap
against intercept at this seed ($p_{\text{corr}}{=}0.012$) did
not reproduce at seed-$2$ ($-3$\,pp, $p_{\text{corr}}{=}1.0$;
pooled $+7.5$\,pp, upper CI ${\approx}15$\,pp), so we read it
as a single-seed draw from the envelope rather than an
established pull-specific perturbation. At trial~1, where
coordination is logically active, all
three protocols collapse to $0.54$.
\textbf{(b) Paired trial-1 outcomes} over $n{=}100$ tasks.  Ties
dominate; all three sign tests fail to reject the no-difference
null ($p{=}1.0$). Informative pairs after ties on the
coordination-active subsets of
Table~\ref{tab:tau2retail:trial1} are $10/8/17$, so this is no
detectable effect at the available power rather than a clean
null.}
\label{fig:headline}
\Description{Two-panel figure. Panel (a) is a grouped bar chart of
success rate (trial 0 vs trial 1) for no_coord, pull, intercept;
trial 1 is flat at 0.54 for all three. Panel (b) is a horizontal
stacked bar chart of wins/losses/ties for the three pairwise
contrasts; ties dominate.}
\end{figure*}

\paragraph{Setup and headline.}
100 tasks (seed 42) $\times$ 3 protocols $\times$ 2 trials gives
600 trials in total.  Both the assistant and the simulated user
are \texttt{claude-haiku-4-5} at $T{=}0$, with a 20-turn budget
per trial.  The task-scoped store is shared across the 2 trials
within a protocol cell.  Figure~\ref{fig:headline} summarizes the
result.  Table~\ref{tab:tau2retail:headline} reports cell-level
metrics; Table~\ref{tab:tau2retail:trial1} reports the
coordination-active paired tests.

\begin{table}[t]
\centering
\caption{$\tau^2$-bench retail headline results
($n=100$ tasks $\times$ 2 trials per protocol, 200 trials per
cell, Claude Haiku~4.5; zero infra errors across 600 trials).
All three protocols collapse to the same trial-1 success rate
($0.540$): coordination does not move the metric once the store
is populated.}
\label{tab:tau2retail:headline}
\small
\begin{tabular}{lrrrr}
\toprule
Protocol & succ.\ rate & pass$^1_{\mathrm{Chen}}$ & pass$^2_{\mathrm{Chen}}$ & Trial-1 succ.\ \\
\midrule
\texttt{no\_coord}  & 0.540 & 0.540 & 0.380 & 0.540 \\
\texttt{pull}       & 0.580 & 0.580 & 0.440 & 0.540 \\
\texttt{intercept}  & 0.490 & 0.490 & 0.330 & 0.540 \\
\bottomrule
\end{tabular}
\flushleft\footnotesize
Marginal cell metrics for the three protocols.
pass$^k_{\mathrm{Chen}}$ follows the Chen et al.\ unbiased
estimator (marginal success rate at $k{=}1$; both-trials-succeed
rate at $k{=}2$).  The \emph{trial-1 succ.\ rate} column reports
the success rate on trial~1 only, where the trace store contains
peer events from trial~0 and coordination is logically active.
\end{table}

\begin{table}[t]
\centering
\caption{Coordination-active pass$^k$ test (P4, defined in \S\ref{sec:eval:metrics};
\S\ref{sec:proposals}): paired sign tests on trial-1 outcomes
restricted to tasks where \emph{both} compared protocols had a
trial-0 failure, so both stores contain peer events and the
coordination mechanism is necessarily active.  Subset $n$ varies
by contrast.  All three pairwise comparisons fail to reject on
this subset; with $10/8/17$ informative pairs after ties the
test is underpowered for a medium-sized effect, so this is no
detectable effect at the available power rather than a clean
null.  The recovery-rate ordering against the \texttt{no\_coord}
baseline shows both coordination architectures slightly
underperforming no coordination at recovering from prior failure
(W/L/T = wins/losses/ties).}
\label{tab:tau2retail:trial1}
\small
\begin{tabular}{lrcrr}
\toprule
Comparison & subset $n$ & t1 succ.\ rate & W/L/T & sign $p$ \\
\midrule
pull vs.\ intercept & 29 & 0.28 / 0.28 & 5 / 5 / 19 & $1.00$ \\
pull vs.\ no\_coord & 28 & 0.14 / 0.29 & 2 / 6 / 20 & $0.29$ \\
intercept vs.\ no\_coord & 35 & 0.20 / 0.34 & 6 / 11 / 18 & $0.33$ \\
\bottomrule
\end{tabular}
\flushleft\footnotesize
Each contrast uses its own paired subset; rows are not
cross-comparable.  The high tie counts ($19$, $20$, $18$) leave
informative samples of $10$, $8$, $17$ tasks, so these sign
tests are near-powerless and the directional recovery deficits
are hypothesis-generating only.  Unconditional trial-1 paired
sign tests over all $n{=}100$ tasks return identical ties
(pull-vs-intercept $13/13/74$; pull-vs-no\_coord $19/19/62$;
intercept-vs-no\_coord $20/20/60$; all $p{=}1.0$).  This table
conditions on trial-0 failure; the pass$^2$ column in
Table~\ref{tab:tau2retail:headline} counts both-trials-success
unconditionally over all 100 tasks.  The two are arithmetically
consistent: pull's pass$^2{=}0.44$ equals $0.71 \times 62/100$,
where $0.71$ is pull's conditional trial-1 success rate given
trial-0 success.
\end{table}

\paragraph{Coordination-active head-to-head: no detectable
effect at this power.}
The trial-1 outcome on the trial-0-failed subset was pre-specified
as the architectural test (P4; \S\ref{sec:eval:metrics}): on those
tasks the store contains peer events, so the coordination
mechanism is active. Pull and intercept tie at trial-1 success
rate $0.28$ on the 29-task both-failed subset (paired sign test
$W/L/T=5/5/19$, $p=1.0$). Each directionally underperforms
\texttt{no\_coord} on its own trial-0-failed subset (pull $0.143$
vs.\ $0.286$, $n{=}28$; intercept $0.20$ vs.\ $0.343$, $n{=}35$),
but neither reaches significance at this subset size. After
ties, the informative pairs number $10/8/17$, well below the
power needed to detect a medium-sized effect; ``null'' would
overclaim, and we read this as no detectable effect at the
available power rather than evidence that the mechanisms are
inert. The unconditional trial-$1$ test over all 100 tasks
returns identical ties. The takeaway is that the naive
deployed-today coordination forms (last-$K$ writers paired with
either curated reading or per-match injection) do not transfer
peer state into pass$^k$ at this scale and model, and the
directional evidence is at least consistent with mild
interference rather than help; both readings would require a
larger paired sample to separate.

\paragraph{Trial-0 decomposition: an empirical noise-floor estimate.}
At trial~0 the store is empty under every protocol, so the
coordination mechanism is inactive. Of the three pairwise
contrasts, only no\_coord-vs-intercept is genuinely
configuration-equivalent: code inspection (see
\texttt{protocols.py}) confirms that both arms reduce to
pure no-ops on the empty store. Pull's reader still re-enters
its augmenter on every trial-$0$ turn (which returns input
unchanged when \texttt{warnings\_block()} is empty), so its
contrast with the other two arms is not configuration-equivalent
in the same sense, so its contrasts must be read separately
from the clean one.
Paired sign tests at seed-$1$ ($n{=}100$, Bonferroni $m{=}3$;
the second seed follows below):
no\_coord-vs-intercept $21/11/68$
($p_{\text{corr}}{=}0.330$); pull-vs-no\_coord $18/10/72$
($p_{\text{corr}}{=}0.555$); pull-vs-intercept $27/9/64$
($p_{\text{corr}}{=}0.012$). Wilson $95\%$ CIs on the signed
gaps are $[{-}2,{+}16]$\,pp, $[{-}1,{+}19]$\,pp, and $[{+}6,{+}26]$\,pp
respectively: the first two cross zero, the third does not---at
this seed.

\paragraph{Reproducibility: a second seed.}
We re-ran the full $n{=}100$ retail sweep at a second seed
(\texttt{seed42$\to$seed-2}, same task IDs, same harness, fresh
API draws). Trial-$1$ rates remain effectively flat
($0.520/0.540/0.510$ for no\_coord/pull/intercept); the
head-to-head still shows no detectable effect. Trial-$0$ paired
sign tests at seed-$2$: no\_coord-vs-intercept $13/13/74$
($0$\,pp signed gap, $p_{\text{corr}}{=}1.0$); pull-vs-no\_coord
$15/18/67$ ($-3$\,pp, $p_{\text{corr}}{=}1.0$);
\textbf{pull-vs-intercept $18/21/61$
($-3$\,pp, $p_{\text{corr}}{=}1.0$)}.  The seed-$1$ $+18$\,pp
pull-vs-intercept significance does not reproduce; the sign
even flips. We correspondingly retract E4 (pull-specific
structural perturbation) as an established finding---a single
seed's $p_{\text{corr}}{=}0.012$ on a contrast that pools to
$45/30/125$ ($+7.5$\,pp, $p_{\text{corr}}{=}0.316$) over both
seeds is exactly the artifact a paired noise-floor protocol
should catch, and ours did.

Pooled across both seeds ($n{=}200$): no\_coord-vs-intercept
$34/24/142$ ($+5$\,pp, CI $[{-}2,{+}12]$\,pp,
$p_{\text{corr}}{=}0.711$); pull-vs-no\_coord $33/28/139$
($+2.5$\,pp, $p_{\text{corr}}{=}1.0$); pull-vs-intercept
$45/30/125$ ($+7.5$\,pp, CI $[{-}1,{+}15]$\,pp,
$p_{\text{corr}}{=}0.316$). \emph{No paired contrast is
significant after Bonferroni at any single seed or pooled.}
The clean-contrast pooled signed gap is $\approx 5$\,pp with
Wilson upper bound $\sim 12$\,pp; the largest pooled upper bound
across any contrast is $\sim 15$\,pp (pull-vs-intercept). The
across-seed observed range of signed gaps is
$[{-}3,{+}18]$\,pp, with the upper end anchored on a single
seed that did not replicate. We therefore characterize the
configuration-equivalent floor as a paired-disagreement
envelope whose pooled upper CI is $\lesssim 15$\,pp; calling
it ``10--18\,pp'' from one seed alone overclaimed the
precision. The motivation for P4 (coordination-active
pass$^k$, \S\ref{sec:eval:metrics}) stands: even a
${\sim}15$\,pp envelope swamps the small headline gains
recent coordination architectures report.

\paragraph{Mechanism diagnosis: three failure modes.}
\emph{(M1) Writer mis-attribution.} The causally-relevant call is
often upstream; last-$K$ records downstream symptoms. A
single-judge Haiku 4.5 LLM-judge audit on $30$ analyzable
trial-$0$ failures returns $77\%$ \textsc{upstream}
(direction-of-magnitude motivation, not a confirmed effect; see
L4 in \S\ref{sec:eval:limits}).
\emph{(M2) Per-match injection noise.} The intercept store
accumulates $\bar{n}{=}2.77$ events per trial (max $8$), and each
matching call fires a separate warning.
\emph{(M3) Brittle keying.} Literal-tuple payloads do not
transfer: \texttt{cancel\_order(\#W123)} does not match
\texttt{cancel\_order(\#W456)} even when both target shipped
orders. \S\ref{sec:proposals} discusses refinements matched
to M1--M3.

\subsection{Cross-model and cross-domain probes}
\label{sec:eval:results:sonnet}
\label{sec:eval:results:airline}

\begin{figure*}[t]
\centering
\includegraphics[width=0.92\textwidth]{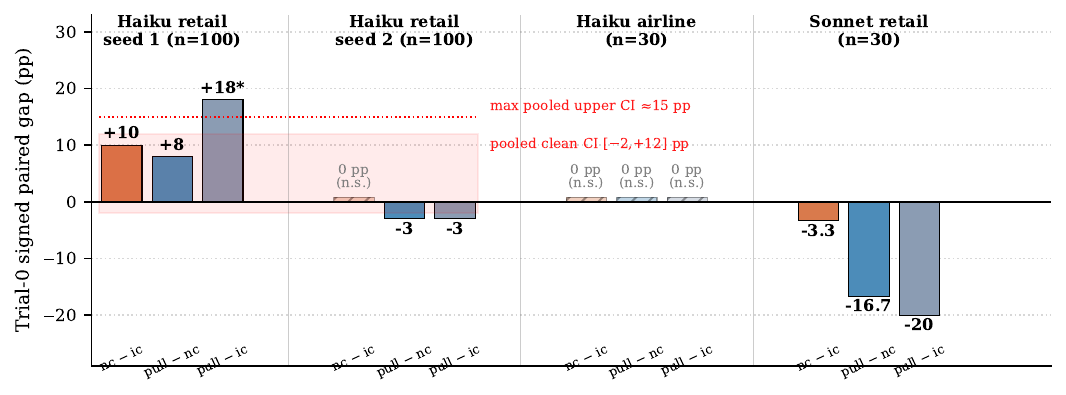}
\caption{Trial-0 paired-sign-test signed gaps across seed,
model, domain, and contrast. The noise floor is a measurement
output of the coordination-active pass$^k$ protocol, not a
universal constant. Haiku retail: the seed-$1$ gaps
($+10/{+}8/{+}18$\,pp) collapse to $0/{-}3/{-}3$\,pp at
seed-$2$; the red band marks the pooled clean-contrast CI
$[{-}2,{+}12]$\,pp and the dotted line the largest pooled upper
Wilson CI ${\approx}15$\,pp. The starred seed-$1$ $+18$\,pp bar
is the non-replicating single-seed result retracted in
\S\ref{sec:eval:results:retail-intercept}. The floor is domain-
and capability-dependent: Haiku airline at $n{=}30$ shows
$0$\,pp gaps across all three contrasts, while Sonnet retail at
$n{=}30$ spans magnitudes $3.3$--$20$\,pp with the direction
flipped---pull starts trial-0 \emph{behind} \texttt{no\_coord}
on Sonnet rather than ahead---so the $20$\,pp magnitude there
sits slightly above the upper edge of the Haiku-retail
observations.}
\label{fig:noise-floor}
\Description{Bar chart of signed trial-0 paired gaps in four
groups. Haiku retail seed 1 at +10/+8/+18pp; Haiku retail seed
2 at 0/-3/-3pp; Haiku airline at 0pp; Sonnet retail at
-3.3/-16.7/-20pp. Red band marks the pooled clean-contrast CI
[-2,+12]pp over the two Haiku retail groups, with a dotted line
at the +15pp largest pooled upper CI.}
\end{figure*}

Two $n{=}30$ probes test whether the underpowered non-result
and the floor are Haiku-retail-specific;
Figure~\ref{fig:noise-floor} summarizes the trial-$0$ gaps.
The probes illustrate the metric rather than making
coordination claims at this scale.

\paragraph{Sonnet 4.5 retail $n{=}30$.}
Trial-1 rates: \texttt{no\_coord} $0.500$, pull $0.733$,
intercept $0.733$; raw pull-vs-no\_coord $p{=}0.039$ does not
survive Bonferroni $m{=}3$. The trial-$0$ gap magnitudes span
$3.3$--$20$\,pp (overlapping the Haiku-retail observations,
with the largest magnitude slightly above their upper edge)
but with the direction flipped: pull starts $-16.7$\,pp
\emph{behind} at trial~$0$, so its $+23$\,pp marginal gain partly
recovers an unrelated trial-$0$ deficit. The coordination-active
metric makes this direction-flip visible; marginal pass$^1$
obscures it. Sonnet $n{\geq}250$ is pre-registered as
confirmatory.

\paragraph{Haiku airline $n{=}30$.}
Trial-1 rates: \texttt{no\_coord} $0.500$, pull $0.533$,
intercept $0.567$ (all paired $p{>}0.69$); trial-0 gaps collapse
to $\approx 0$\,pp, so the floor is domain-dependent. On the
coordination-active subset ($n{=}14$), the direction matches
retail (\texttt{no\_coord} $0.357$, intercept $0.286$, pull
$0.214$): the naive last-$K$ scheme does not help and plausibly
impairs recovery.

\subsection{From noise floor to deployment threshold}
\label{sec:eval:deployment}

Converted to a release-time decision rule, the Haiku-retail
envelope (pooled upper Wilson CI ${\approx}15$\,pp,
\S\ref{sec:eval:results:retail-intercept}) becomes a
\emph{minimum detectable effect (MDE)} for sample-size planning
within that regime. Because the floor varies by domain (Haiku
airline $\approx 0$\,pp) and capability tier (Sonnet retail
magnitudes $3.3$--$20$\,pp), the MDE table is regime-specific, and transport
to other model/domain pairs requires re-running the paired
protocol. Applying a two-proportion $z$-test at $\alpha{=}0.05$,
power $0.80$, baseline $p_1{=}0.50$, Table~\ref{tab:mde} gives
the required sample sizes.

\begin{table}[t]
\centering
\caption{Minimum sample size (tasks per arm) to detect a
candidate pass$^k$ improvement $\Delta$ on $\tau^2$-bench retail
at $\alpha{=}0.05$, power $0.80$, baseline $p_1{=}0.50$.
``Indep.'' is the two-proportion $z$-test for independent arms.
``Paired'' is the McNemar test calibrated on the seed-$1$
trial-0 discordance rate ($p_d{=}0.32$, \texttt{no\_coord} vs.\
intercept; pooled across both seeds $p_d{=}0.29$, which barely
moves the column).  Read-out: $\tau^2$-bench retail's public
114 tasks at one seed detect only $\Delta{\geq}20$\,pp at the
independent-arm rate ($n{=}93$), or $\Delta{\geq}15$\,pp
paired ($n{=}110$).}
\label{tab:mde}
\small
\begin{tabular}{rrrr}
\toprule
$\Delta$ (pp) & Indep.\ $n$ & Paired $n$ & $\tau^2$-retail seeds$^\ast$ \\
\midrule
5  & 1565 & 1003 & impractical \\
10 & 388  & 249  & 4 seeds / 3 paired \\
\textbf{15} & \textbf{170}  & \textbf{110}  & \textbf{2 seeds / 1 paired} \\
18 & 117  & 76   & 2 seeds / 1 paired \\
20 & 93   & 61   & 1 seed / 1 paired \\
25 & 58   & 38   & 1 seed / 1 paired \\
\bottomrule
\end{tabular}
\flushleft\footnotesize
$^\ast$ Number of full $114$-task seeds required to reach $n$;
``paired'' counts the paired condition.
\end{table}

\paragraph{Motivation only: literature gap-size context.}
Table~\ref{tab:literature-vs-floor} sets ten recent multi-agent
coordination contributions next to the Haiku-retail pooled
envelope: seven headline gains fall below the clean-contrast
pooled upper CI ($12$\,pp) and one more sits inside the
$12$--$15$\,pp band up to the largest pooled upper CI. The ten are recent (2023--2026) multi-agent
coordination architecture papers we found citing
MAST~\cite{mast2025} or CodeDelegator~\cite{codedelegator2026}
and reporting a single-number headline architectural delta on a
named benchmark; this is illustrative rather than systematic.
The point is to motivate reporting same-model paired floors
alongside future coordination claims, not to adjudicate the
underlying systems in their original (different model, task,
metric, $n$) settings, which would require per-row paired
replication. We acknowledge the tension head-on: this paper
argues against single-number cross-setting comparisons, and
Table~\ref{tab:literature-vs-floor} is precisely that
juxtaposition; we include it as a motivating prompt because
the gap-size visibility outweighs the methodological cost,
not because the comparison itself adjudicates anything.

\begin{table}[t]
\centering
\caption{Motivation examples for why same-model paired floors
matter; not a verdict on the underlying systems. Each row's
effect is shown alongside our Haiku-retail pooled envelope
purely to motivate the protocol; the comparison is heterogeneous
(different model, task, metric, $n$) and any individual row may
be entirely real in its original setting. ``vs.\ floor'' marks
whether the published gap is below the clean-contrast pooled
upper CI of $12$\,pp ($\downarrow$), inside the $12$--$15$\,pp
band up to the largest pooled upper CI ($=$), or above
$15$\,pp ($\uparrow$); the floor itself varies by model
and domain (\S\ref{sec:eval:results:airline}).}
\label{tab:literature-vs-floor}
\small
\setlength{\tabcolsep}{4pt}
\begin{tabular}{lll@{\hspace{6pt}}c}
\toprule
Paper & Bench / Task & Effect & vs.\ floor \\
\midrule
CA-MCP~\cite{camcp2026}            & TravelPlanner BERTScore & $+1.2$ pts & $\downarrow$ \\
Terrarium~\cite{terrarium2025}     & Meeting Scheduling      & $1.0$--$2.7$\,pp & $\downarrow$ \\
AgentVerse~\cite{agentverse2023}   & HumanEval (Solo vs.\ Group)  & $+1.8$\,pp  & $\downarrow$ \\
AutoGen~\cite{autogen2023}         & MathChat over PoT/PS    & $\sim$$+6$\,pp & $\downarrow$ \\
MetaGPT~\cite{metagpt2024}         & SoftwareDev quality     & $<10$\,pp & $\downarrow$ \\
Silo-Bench~\cite{silobench2026}    & GPT-OSS P2P vs.\ BP     & $+6$\,pp & $\downarrow$ \\
Reflexion~\cite{reflexion2023}     & HumanEval                  & $+11$\,pp   & $\downarrow$ \\
\midrule
Ou et al.~\cite{ou2025coordination} & TravelPlanner orch.+notebook & $+13.5$\,pp & $=$ \\
\midrule
Reflexion~\cite{reflexion2023}     & ALFWorld                    & $+22$\,pp & $\uparrow$ \\
Ou et al.~\cite{ou2025coordination} & TravelPlanner combined     & $+17.5$\,pp & $\uparrow$ \\
\bottomrule
\end{tabular}
\end{table}

Coordination-active pass$^k$ is the minimum pre-deployment
measurement gate that any coordination claim on $\tau^2$-bench
(and the broader $\tau$-bench family of state-validated
benchmarks) should pass; Table~\ref{tab:mde} ships the
release-gate.

\paragraph{Monitoring schema.}
Three runtime rules for AgentCore / SageMaker Model
Monitor~\cite{sagemaker_monitor2022} deployments:
(R1)~CUSUM on coordination-active pass$^k$ over a golden-task
set, alarm at $\pm 1.5\sigma_{\textrm{floor}}$;
(R2)~rolling-$N$ $95$th-percentile injection-event count, alarm
at $\geq 50\%$ above release baseline (right-skew defeats
Gaussian $2\sigma$);
(R3)~coordination-key collision rate, alarm at $>0.5\%$.

\paragraph{Relation to release-gate and reliability work.}
The Anthropic infrastructure-noise
study~\cite{anthropicnoise2026} reports $6$\,pp configuration
gaps on SWE-Bench from CPU/RAM in a single-agent setting;
Mustahsan~et~al.~\cite{mustahsan2025stochasticity} propose an
intraclass correlation coefficient (ICC) framework on
FRAMES/GAIA; CI/CD quality
gates~\cite{selftestquality2026} and behavioral drift
monitoring~\cite{agentdrift2026} cover adjacent ground. Ours is
the first MDE-grounded release-gate keyed on a
coordination-architecture noise floor with store-conditional
restriction on a state-validated coordination benchmark.

\subsection{Limitations}
\label{sec:eval:limits}

Six limitations bound the claims above.
\textbf{L1 (identity-replay):} request-equivalence is verified by
code inspection plus a $40/40$ SHA-$256$ payload-hash audit on a
separate $n{=}10$ retail sample ($10$ tasks $\times$ $4$
first-request payload components, each byte-identical across
all three protocols;
\S\ref{sec:eval:request-equivalence}); the headline $n{=}100$
sweep was not itself byte-audited, but the audit shows the
request-assembly invariant holds. The Messages API exposes no
seed, so the pooled envelope (upper CI ${\approx}15$\,pp)
upper-bounds the residual server-side stochasticity floor.
\textbf{L2 (writer pilot-tuning):} the writer ($K{=}3$, single
\textsc{failed\_trial\_action} type) was tuned on a prior $n{=}30$
sweep; $\textsc{pass}^k_{\mathrm{active}}$ is pre-specifiable in
principle but was applied to already-collected data here.
\textbf{L3 (within-protocol replicate + second seed).} A paired
$n{=}30$ within-protocol audit gives trial-$0$ signed gaps of
$0$--$3.3$\,pp across all three protocols ($p{=}1.0$), bounding
pure API stochasticity at $\leq 3$\,pp. A second
$n{=}100$ headline seed
(\S\ref{sec:eval:results:retail-intercept}) confirms that no
single-seed paired contrast is significant once we measure
twice: pooled across both seeds, all three trial-$0$ contrasts
land at signed gaps of $2.5$--$7.5$\,pp with Wilson upper CIs
$\leq 15$\,pp and Bonferroni $p_{\text{corr}} > 0.3$. Pull's
seed-$1$ $+18$\,pp gap against intercept was not reproduced
($-3$\,pp at seed-$2$), so we retract E4 (pull-specific
perturbation) as an established source.
\textbf{L4 (single-judge M1):} the $77\%$ upstream rate uses a
same-model-class judge with no Cohen's $\kappa$ or independent
replication; M1 is a hypothesis for confirmatory work.
\textbf{L5 (cross-model/-domain $n{=}30$):} Sonnet retail and
Haiku airline show regime-specificity rather than coordination
claims; Sonnet $n{\geq}250$ is pre-registered. The $n{=}100$
headline's coordination-active subsets are tight (pull vs.\
intercept paired, $n{=}29$, $5/5/19$), so the architectural
head-to-head is an underpowered non-result rather than a clean
null, and is weaker evidence than the measurement contribution.
\textbf{L6 (trial-$0$ inertness):} the protocol assumes
peer-coordination flows reader-side only; writer-side anticipation
effects are out of scope.
\textbf{Scope.} Haiku $4.5$ retail carries the $n{=}100$
headline; outcome conversion is not detectable at the available
power at this model and scale.

\section{Forward Work}
\label{sec:proposals}
\label{sec:discussion}

The coordination-active null
(\S\ref{sec:eval:results:retail-intercept}) isolated three failure
modes of naive coordination, each with a matched refinement on
existing production hook surfaces (Strands~\cite{strandshooks2025},
AgentCore Policy~\cite{agentcorepolicy2025}, LangChain
middleware~\cite{langchainmiddleware2026}).
\textbf{P1: causally-attributed writers} address \emph{M1
writer mis-attribution}: an LLM-judge consumes the trajectory
and reward and re-weights which trial calls get written, replacing
the recency-only last-$K$ heuristic.
\textbf{P2: selective intercept} addresses \emph{M2 per-match
injection noise}: a score gate fires only when a peer event is
relevant, novel, and decision-proximal (preliminary signal below).
\textbf{P3: predicate constraint extraction} addresses \emph{M3
brittle literal keying}: an extractor turns the trajectory into a
typed predicate (e.g.\ \texttt{tool=cancel\_order} with
\texttt{order.status=shipped}) so intercept fires on the
predicate, not the literal argument tuple.
\textbf{P4: coordination-active pass$^k$}
(\S\ref{sec:eval:metrics}) is the release-time gate any refinement
should clear. P1 and P3 each require an LLM-judge component not
validated here. A shared trace store also opens security surfaces
(poisoned writes, cross-principal exfiltration); the prototype
assumes a trusted single-tenant deployment.

\paragraph{P2 (selective intercept), preliminary signal.}
A score-gated intercept that fires only when a peer event is
relevant, novel, and decision-proximal beat vanilla intercept on
the coordination-active subset at Haiku~4.5 retail $n{=}30$
($+15.8$\,pp, paired sign $2/7/21$, $p{=}0.18$; hyperparameters
not pre-registered). Directionally consistent and underpowered;
Sonnet at $n{\geq}250$ is the natural confirmatory test.

\section{Conclusion}
\label{sec:conclusion}

The observed trial-0 paired-gap envelope on $\tau^2$-bench
retail spans $[{-}3,{+}18]$\,pp across all three contrasts and
two $n{=}100$ Haiku 4.5 seeds; the clean
configuration-equivalent contrast (no\_coord vs.\ intercept)
gives $+10$ and $0$\,pp, pooling to $+5$\,pp (CI
$[{-}2,{+}12]$), with a pooled upper Wilson CI of
${\sim}15$\,pp on the largest between-protocol contrast; no
single-seed Bonferroni significance survives a second seed. ET-MCP, the task-scoped
trace-store substrate, made the matched-payload comparison
possible; the measurement discipline, not the substrate, is
what we expect to travel. Notably, our own seed-$1$
pull-vs-intercept $p_{\text{corr}}{=}0.012$ did not reproduce
at seed-$2$---a worked example of the protocol catching the
single-seed artifact it is designed to catch.

Seven of ten recent coordination architectures report headline
gains below that envelope and one more sits inside it. The
coordination-active head-to-head here is an underpowered
non-result, not a clean null: the informative pairs after ties
are $n{=}8$--$17$, well short of what a medium-sized effect
would require, so it reframes the question rather than retiring
it. The negative reading applies to \emph{naive} coordination
under a measurement-honest protocol, not to coordination as a
research program. P1--P3 (\S\ref{sec:proposals}) are the
mechanisms we expect to clear the floor, and
coordination-active pass$^k$ is the gate they should be judged
against. Reporting a same-model paired floor alongside any
multi-agent delta is the practice the field can adopt
now, with the protocol here as a starting template.

\bibliographystyle{ACM-Reference-Format}
\bibliography{references}

\end{document}